\documentclass[aps,pre,preprint,nofootinbib,groupedaddress,showpacs,showkeys]{revtex4}
\usepackage{epsfig}

\begin{document}
\preprint{APS preprint}
\title{Stochastic Opinion Formation in Scale-Free Networks}

\author{M. Bartolozzi$^{1}$,
D. B. Leinweber$^{1}$, A. W. Thomas$^{2,1}$}
\affiliation{$^1$Special Research Centre for the Subatomic 
Structure of Matter (CSSM),
 University of Adelaide, Adelaide, SA 5005, Australia\\
$^2$Jefferson Laboratory, 12000 Jefferson Ave., Newport News, VA 23606, USA}
\date{\today}

\begin{abstract}
The dynamics of opinion formation in large groups of people is a
complex non-linear phenomenon whose investigation is just at the
beginning.  Both collective behaviour and personal view play an
important role in this mechanism.  In the present work we mimic the
dynamics of opinion formation of a group of agents, represented by two
states $\pm 1$, as a stochastic response of each agent to the
opinion of his/her neighbours in the social network and to feedback
from the average opinion of the whole.  In the light of recent studies, a
scale-free Barab$\acute{a}$si-Albert network has been selected to
simulate the topology of the interactions.  A turbulent-like dynamics,
characterized by an intermittent behaviour, is observed for a certain
range of the model parameters.  The problem of uncertainty in decision
taking is also addressed  both from a topological point of view, using
random and  targeted removal of agents from the network, and by  
implementing a three state model, where the third state, zero, is
related to the information available to each agent.  Finally, the
results of the model are tested against the best known network of
social interactions: the stock market. A time series of  daily
closures of the Dow Jones index has been used as an indicator  of the
possible applicability of our model in the financial context.  Good
qualitative agreement is found.
\end{abstract}

%\end{keyword}                                   
%\end{frontmatter}
\maketitle
% main text
\section{Introduction}
\label{}

Systems composed of many parts that interact with each other  in a
non-trivial way are often referred to as {\em complex systems}. The
social relations between individuals can perhaps be included in this
category.  An intriguing issue concerns the role played by the
topological structure of the social network in governing the dynamical 
behaviour of the system.

Recent studies of the topological properties of interactions in
 different biological,  social and technological systems has made
 it possible to shed some light  on the basic principles of structural
 self-organization.  A few  examples include the food
 webs~\cite{food_web}, power grids and neural
 networks~\cite{watts98,amaral00},  cellular networks~\cite{cellular},
 sexual contacts~\cite{sexual},  Internet routers~\cite{internet}, the
 World Wide Web~\cite{www},  actor
 collaborations~\cite{watts98,albert99,amaral00,act}, the citation
 networks of scientists~\cite{citations} and the stock
 market~\cite{market}.  Although different in the underlying
 interaction dynamics or {\em micro-physics}, all these
 networks have shown a tendency to self-organize in structures that
 share common features. In particular, the number of connections, $k$,  for
 each element, or node, of the network follow a  power law
 distribution, $P(k)\sim k^{-\alpha}$. Networks that fulfill this
 property  are referred to as {\em scale-free} (SF) networks.
 In addition many of these networks are characterized by a high
 clustering coefficient, $C$, in comparison with random
 graphs~\cite{bollobas}. The clustering coefficient, $C$, is computed as the
 average of local clustering, $C_i$, for the $i$th node, defined as
\begin{equation}
 C_i=\frac{2y_i}{z_i(z_i-1)},
\end{equation} 
where $z_i$ is the total number of nodes linked to the site $i$ and $y_i$  is
the total number of links between those nodes.  As a consequence
both  $C_i$ and $C$ lie in the interval [0,1].
The high level of clustering found
supports the idea that a {\em herding} phenomenon is a common feature
 in social and biological communities.

Numerical studies on SF networks have demonstrated how the topology
plays a fundamental role in infection spreading~\cite{pastor01} and
tolerance against random and preferential node
removal~\cite{tolerance}.  A detailed description of the progress in
this emerging field of statistical mechanics can be found in the
recent reviews of Refs.~\cite{albert02,dorogovtsev02}.  In the present
work we investigate the implication of a scale-free topology  in a
stochastic opinion formation model. Similar versions of this model
have been tested in regular  lattices~\cite{kaizoji,krawiecki02} and
percolation clusters~\cite{bartolozzi04}. These models adopt a
mean field approach where the interactions are extended between all
the individuals in the lattice or cluster respectively. In contrast, our
simulation focuses on the role of short-range first-neighbour interactions
for cases where the topological structure of the interactions is not trivial.

In the next section we describe the model used for the simulation. In
Sec.~\ref{numerical} we  show the results obtained  numerically while 
in Sec.~\ref{indecision} we investigate the importance
of failures in the network during the process of opinion formation
while Sec.~\ref{3states} the two state model is  extended to  three
states and the numerical results are compared. In Sec.~\ref{dj}
and~\ref{mf_analysis} we test the results of our  simulations against
the best known social network: the stock market.  In particular the
time series of average opinion fluctuations obtained with the  model
is compared with the time series of price variations for the Dow Jones
index from 13/1/1930 to 13/4/2004. The final section presents further
discussion and conclusions.

\section{The Model}

In the present work we investigate the opinion formation process of a
group of $N$ individuals, represented by nodes on a SF network.  The
mechanism of opinion formation is simulated using  stochastic heat
bath dynamics with feedback.  The opinion of each agent  is of a
Boolean type. That is, at each discrete time step, $t$, the opinion is
represented  by one of two possible states (or spin orientations),
namely $\sigma_{i}(t)= \pm 1$, for the $i$th agent.  A practical
example could be the decision to buy, $\sigma_{i}(t)=+1$, or sell,
$\sigma_{i}(t)=-1$, a stock in a virtual stock market.
 
In order to mimic the scale-free network topology we make use  of the
Barab$\acute{a}$si-Albert model~\cite{albert99}. This is based on two
main assumption: (i) linear growth and (ii) preferential attachment.
In practice the network is initialized with $m_0$ disconnected
nodes. At each step a new node with $m$ edges is added to the
pre-existing network.  The probability that an edge of the new node is
linked with the $i$th node is expressed by $\Pi(k_i)=k_i/\sum_{j}k_j$.
The iteration of this preferential growing process
yields a scale free network, $P(k)\sim k^{-\alpha}$ where $\alpha=3$.

\begin{figure}
\centerline{\epsfig{figure=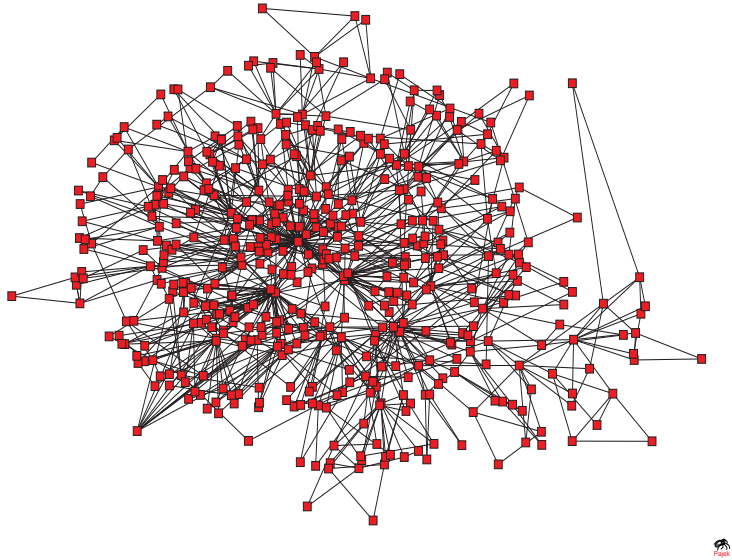,height=8cm, width=10cm}}
\caption{ (Color online). Example of a scale-free network. The number of nodes
 is 500 with clustering probability $\theta=0.9$ and $m_0=m=2$, so that each
new node is linked twice.
 The number of nodes has been
kept small in order to preserve the clarity of the plot. Note that, for 
such small networks, a large scale invariant range is obtained only 
if one considers the ensemble average over several realizations.
This plot has been realized with the Pajek software~\cite{pajek}.}
\label{netplot}
\end{figure}

It is worth noting that the Barab$\acute{a}$si-Albert model  cannot
reproduce a high clustering coefficient. In fact, the value  of this
coefficient depends on the total number of nodes in the
network~\cite{albert02} and in the thermodynamic limit, $N \rightarrow
\infty$, $C  \rightarrow 0$.  In principle the observed local
clustering can play an important role in the opinion formation of
groups of people, independent of their total number. In order to
account for this, we introduce a further step in the growth process,
namely the triad formation proposed by Holme and Kim~\cite{holme02}. In
this case, if the new added node is linked with an older node, $i$,
having other links, then with a certain probability, $\theta$, the
next link of the new node, if any remain, will be added to a randomly
selected neighbour of node $i$.  This method of introducing friends to
friends, while preserving the scale-free nature of the networks,
generates high clustering coefficients that do not depend on the
number of nodes in the network. The only tunable parameter  that
changes the value of the clustering  coefficient is the 
clustering probability
$\theta$. An example of SF network generated with this 
algorithm is shown in Fig.~\ref{netplot} for 500 nodes \footnote{Another model
for acquaintance networks, showing properties similar to the one
presented in this work, has been proposed by Davidsen {\em et al.}~\cite{davidsen02}.}.

%--------------------------------Figure 1--------------------------------------
%------------------------------------------------------------------------------
 
Once the scale-free network has been built, we randomly assign   the
spin values, $\pm 1$, to  every node and  start the  simulation of opinion
formation. We neglect, in the first approximation,  the network
dynamics. This is equivalent to assuming that the time scale for
evolving the network is much longer that the time needed for people to
make a decision.

The dynamics of the spins follows a stochastic process
that mimics the human uncertainty in decision
making~\cite{krawiecki02,bartolozzi04}.  Values are
updated synchronously according to a local probabilistic rule:
$\sigma_{i}(t+1)=+1$ with probability $p_{i}$ and $\sigma_{i}(t+1)=-1$
with probability $1-p_{i}$.  The probability $p_{i}$ is determined, by
analogy with heat bath dynamics with formal temperature $k_{b}T=1$, by
\begin{equation}
 p_{i}(t)=\frac{1}{1+e^{-2I_{i}(t)}},
\label{heat_bath}
\end{equation}
where the local field, $I_{i}(t)$, is
\begin{equation}
I_{i}(t)=a \xi(t)\tilde{N_i}^{-1}\sum_{j=1}^{\tilde{N_i}}\sigma_{j}(t)
+ h_i \eta_{i}(t) r(t).
\label{field}
\end{equation}
The first term on the right-hand side of Eq. (\ref{field}) represents
the  time dependent interaction strengths between  the node $i$ and
his/her $\tilde{N_i}$  information sources, which are the first
neighbours in the network.  The  second term  instead reflects the
personal reaction to the system feedback, that is the average opinion,
\begin{equation}
r(t)=\frac{1}{N}\sum_{j=1}^N \sigma_j(t),
\end{equation}
 resulting from the previous time step.  The terms $\xi(t)$ and
$\eta_{i}(t)$ are random variables uniformly  distributed in the
interval (-1,1) with no correlation in time nor in the network. They
represent the conviction, at time $t$,
 with which agent $i$ responds to his/her group (common for all the agents)
 and the global opinion of the network respectively.
The strength term,  $a$, is  constant and common for the whole network,
while  $h_i$ is specifically chosen for every individual from a
uniform distribution in (0,$\kappa$) and are both constant  in the
dynamics of the system. By varying the parameter $\kappa$ we can give
more or less weight to the role of feedback in the model.  The
strength coefficients $a$ and $h_i$ in the local field, $I_i$,
characterizing the attributes of the agents, play a key role in the
dynamics of the model.  They represent the relative importance that
each agent of the network gives, respectively, to his/her group and
to the variation of the average opinion itself.

%While $a$ is a parameter associated with the network, $h_i$ is specifically
%chosen for each individual. 

\section{Numerical Simulations}
\label{numerical}

\begin{figure}
\centerline{\epsfig{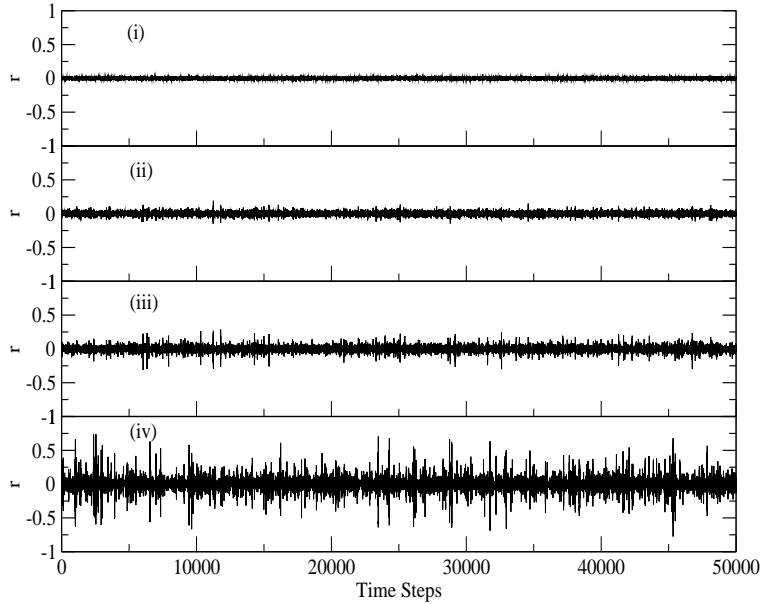}}
\caption{ Time series of the average opinion, $r$,  for different
values of the group interaction strength parameter
 $a$: (i) $a=0.8$, (ii) $a=1.5$, (iii) $a=1.8$
and (iv) $a=2.3$. The parameters used for the simulations are: $N=10^4$
nodes, clustering probability  $\theta=0.9$, initial nodes and links 
per new node $m_0=m=5$
 and we take the upper bound of the distribution of personal response strengths equal
to the group interaction strength, that is $\kappa=a$. The results
involve 10 realizations of the scale free network each displayed for
5000 time steps.  For values of $a$ greater than 1 a turbulent-like
state, characterized by large fluctuations, starts to appear in the
process of opinion formation.  The clustering probability
$\theta=0.9$, related to the triad formation in the network, fixes the
clustering coefficient to $C \approx 0.39$. This value is similar  to
that found for many real systems~\cite{albert02,dorogovtsev02}.}
\label{ts_a}
\end{figure}

At first we investigate the importance of the group strength $a$ 
by fixing $\kappa=a$.  
In this case the dynamical behaviour is similar to that found 
in the stock market context in Refs.~\cite{kaizoji,krawiecki02,bartolozzi04}. 
For $a \alt 1$  the resulting time series of average opinion is largely
uncorrelated Gaussian noise with no particularly 
interesting features, as illustrated in Fig.~\ref{ts_a}(i).

%--------------------------------Figure 2--------------------------------------
%------------------------------------------------------------------------------

As soon as we exceed the value of $a \approx 1$ a turbulent-like
 regime sets in, characterized by large intermittent fluctuations, as
 illustrated in Fig.~\ref{ts_a}(ii $\rightarrow$ iv).  These large
 fluctuations, or   {\em coherent events}, can be interpreted in terms
 of a multiplicative stochastic process with a weak additive noise
 background~\cite{nakao97,krawiecki02}.  For $a > 2.7 $ we observe
 that the bursts of the time series begin to  saturate the bounds $-1
 \le r \le 1$.

\begin{figure}
\centerline{\epsfig{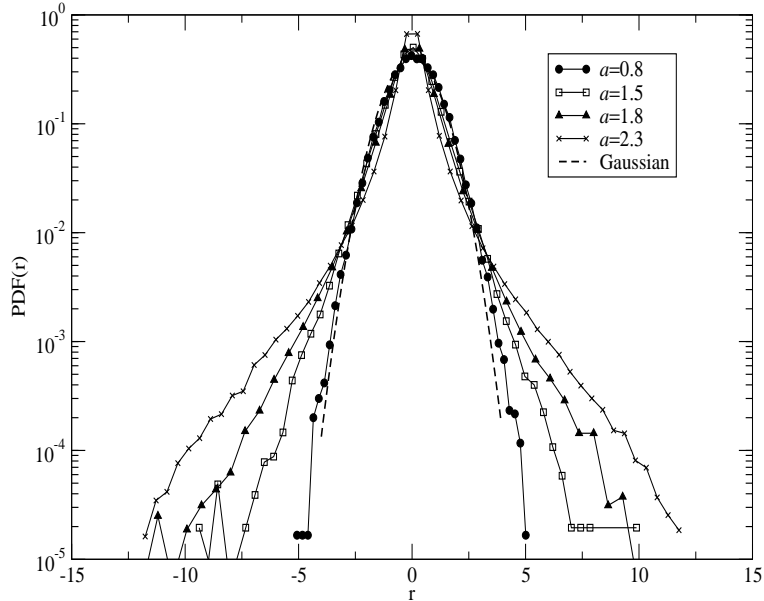}}
\caption{ PDFs of the time series relative to Fig.~\ref{ts_a}. The shapes of
the distributions converge to a Gaussian for small values of the group interaction 
strength $a=\kappa$. A Gaussian 
distribution is also plotted for comparison. All the PDFs in this paper are
obtained over 50 realizations of the SF network.
In order to compare the fluctuations at different scales, 
the time series in the plot have been normalized according to 
$r(t) \rightarrow \frac{r(t)-\bar{r}}{\sigma}$, where $\bar{r}$ and $\sigma$ 
denote the average and the standard deviation over the period considered respectively.}
\label{pdf_a}
\end{figure}

In Fig.~\ref{pdf_a} we plot the probability distribution functions (PDFs) 
Associated with the time series of Fig.~\ref{ts_a}. The large fluctuations, 
for $a$ greater than $\approx 1$, are reflected in the fat tails of the relative PDFs.
Decreasing the value of $a$, and so the number of coherent events,
 the PDF converges to a Gaussian distribution generated by a
 random Poisson process.

%--------------------------------Figure 3--------------------------------------
%------------------------------------------------------------------------------

%If we consider the mean field approximation used in Ref.~\cite{krawiecki02},
%we can notice how the dynamical behaviour of the heat bath systems is similar,
%in a way, to the {\em on-off intermittency} 
%observed in chaotic systems after a {\em blowout
%bifurcation}~\cite{platt93,ott94}.  In these models an additive
%noise, that is intrinsic in the heat bath fluctuations  for our model,
%can enhance a {\em bubbling} regime,  qualitatively similar to the on-off
%%intermittency, in a wide range of the parameter space 
%before the blowout transition~\cite{ashwin94,platt94,ashwin97}.

\begin{figure}
\centerline{\epsfig{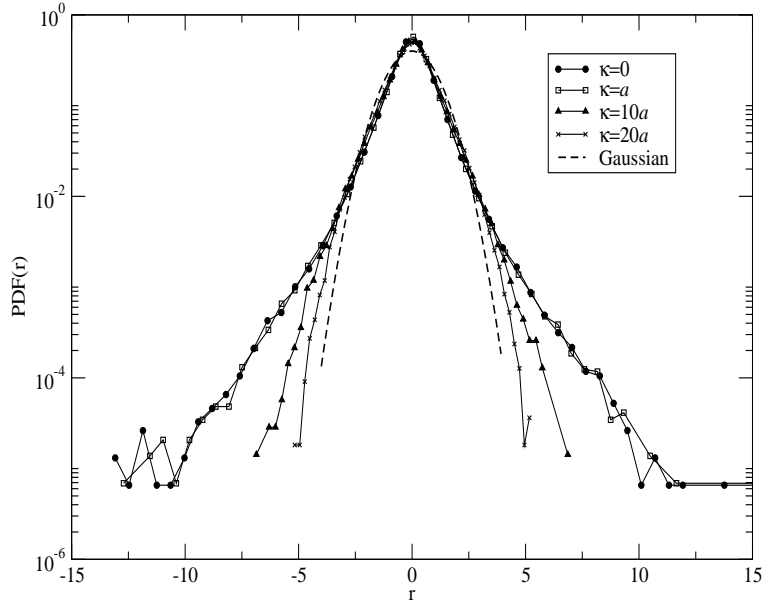}}
\caption{ The importance of the personal response, related to the
 global opinion strength parameter $\kappa$ is shown by the change of shapes of the PDFs
 for group interaction strength $a=1.8$. For large values of $\kappa$ the time series of
 global opinion approaches Gaussian noise. The time series of $r$ has
been normalized -- see the caption of Fig.~\ref{pdf_a}.}
\label{pdf_k}
\end{figure}

The personal response to the change in the average opinion also plays
an important role in the turbulent-like regime of the simulation.  In
order to study the impact of this term on the dynamics we change the
parameter $\kappa$ while keeping $a$ fixed at 1.8. 
The results are summarized by the
PDF plots in Fig.~\ref{pdf_k}.  For $\kappa=0$ the behaviour of the time
series is still turbulent-like, underlying how the network group interaction 
is, in reality, the only crucial factor for the appearance of 
coherent events.  As expected, incrementing the value of $\kappa$ leads to 
a progressive crossover toward a noise regime. It is important to
notice how this regime is  reached for $\kappa > 10a$. 
The group interactions continue to play  an 
important role even when the average value of $h_i$ is large compared to $a$.

%--------------------------------Figure 4--------------------------------------
%------------------------------------------------------------------------------

%The previous simulations have also been repeated for different values
%of $N$, $\theta$ and $m$. While varying the absolute number of agents, $N$,
%and  the average local clustering, $\theta$, does not lead to any substantial
%difference in  the dynamics of the model, the average number of links
%per node, $\bar{k}=2m$, has an effect in the turbulent-like
%phase.  An increase in the average number of links per node
%gives rise to  more turbulence characterized by larger fluctuations and broader
%tails in the PDF.  Large scale synchronizations are more
%likely to occur.  This is just another confirmation of the importance of
%the network group interactions in the formation of  collective opinion.

\begin{figure}
\vspace{1cm}
\centerline{\epsfig{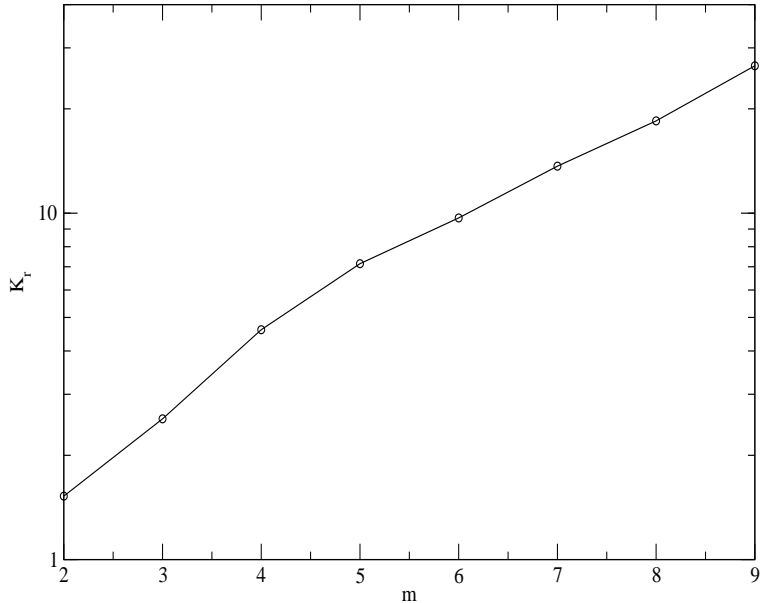}}
\caption{Dependence of the  kurtosis, 
defined as $K_r=\langle r^4 \rangle/ \langle r^2 \rangle^2$, where 
$\langle \dots \rangle$ denotes the temporal average, as a function 
of the node-edge parameter $m$. 
For a Gaussian noise process $K_r=3$ while for $K_r>3$ large deviations
from the average start to appear. The final value for
each $m$ has been obtain after the average over 50 configurations of the network.
The calculations show an exponential increase for $K_r$.}
\label{kurtosis_m}
\end{figure}

In order to test the relevance of the network structure on
the process of opinion formation, the previous simulations have been
repeated, with a large number of nodes, $N$, and $\kappa=a$,  for different values
of the clustering parameter, $\theta$, and the node-edge parameter, $m$. 
While varying $\theta$,  does not lead to any
substantial difference in  the dynamics of the model,  the increase of
the average number of links per node, $\bar{k}=2m$, has a dramatic
effect in the turbulent-like phase, as shown in Fig.~\ref{kurtosis_m}. 
%
%--------------------------------Figure 5--------------------------------------
%------------------------------------------------------------------------------
%
Here the kurtosis, ($K_r=\langle r^4 \rangle/ \langle r^2 \rangle^2$, where 
$\langle \dots \rangle$ denotes the temporal average), of the time series of the
average opinion, used to quantify the deviation from a noise regime,
 is plotted against {\em m}.  It is evident that an
increase in the average number of links per node gives rise to  more
turbulence characterized by larger fluctuations and broader tails in
the PDF. Large scale synchronizations are more likely to occur for large $m$.
This behaviour is intrinsically related to the model of
Eqs.~(\ref{heat_bath}) and  (\ref{field}). In fact, the
turbulent-like regime is a consequence of the random fluctuations of
the interaction strengths between agents  around a 
bifurcation value separating the ordered and disordered phase. 

 If we take
the thermodynamic limit, where $N \rightarrow \infty$ and $m
\rightarrow \infty$ , then the coupling strengths between agents can
be approximated well by the average strength over all the network
 and a mean field approach becomes appropriate to describe the dynamics of the model.
Krawiecki {\em et al.}~\cite{krawiecki02} proposed the following map
\begin{equation}
r(t+1)=A\xi(t)r(t)+h\eta(t),
\label{mean_field_map}
\end{equation}
as a mean field approximation of a stochastic dynamical system similar 
to the one used in the present work. Here $A$ and $h$ are coupling
coefficients and  $\xi(t)$ and $\eta(t)$ random numbers in the
interval (-1,1).  The map of Eq.~(\ref{mean_field_map}) is a generic
model for {\em on-off intermittency} and {\em attractor bubbling}
extensively studied in chaos
theory~\cite{platt93,ashwin94,platt94,ott94,ashwin97}.

It is also worth pointing out that an increase of $\bar{k}$ is 
related to a decrease in the average path length between nodes; that is, the
network ``shrinks'' and becomes more compact. In relation to our
previous discussion, the more compact the network is the more the
dynamics of our system approaches to the mean field approximation.
It becomes easier for the agents to synchronize.  This characteristic of
compactness, referred to as the {\em small world
effect}~\cite{bollobas,albert02,dorogovtsev02}, is actually very common in both
real and artificial networks.

We further investigate the importance of the SF network topology and the 
the small world effect 
in our model by performing a numerical simulation of the same system but using
a {\em random network} (RN) or {\em random graph} as the underlying topology.
  Given a fixed a number of nodes, {\em N}, a RN is defined by the probability $p$ 
that two nodes are linked together~\cite{bollobas,albert02,dorogovtsev02}.  
 In this case $\bar{k}=pN$ and, moreover,  there exists
a critical value, $p\equiv p_c \simeq 1/N$, for which the the network
 undergoes a topological
phase transition where it moves from a phase where  it is
composed of a collection of small, disjoint, sub-networks  to a phase
where a giant cluster emerges \footnote{Note the
analogy between the random network theory and the standard percolation
theory on a lattice~\cite{stauffer92} where the structural properties
of the system are studied as a function a percolation probability.}.
Random networks, while preserving small world properties, have a
Poisson degree distribution~\cite{albert02}, $P(k)$, and small
clustering coefficients. As previously mentioned, we make use of the RN
to test the
robustness of our model  with respect to the topology used and to learn
about the most important properties relevant to the dynamics. 
In order to do so, we fix the number of agents and the average number of links 
for the SF and RN, namely $N=10^4$ and $\bar{k}=10$. 
Then we perform independent numerical simulations on the two topologies. 
Note that for the RN,  $\bar{k}=10$  requires
$p=10/N$, that is ten times greater than the percolation threshold. 
%
%--------------------------------Figure 6--------------------------------------
\begin{figure}
\vspace{1cm}
\centerline{\epsfig{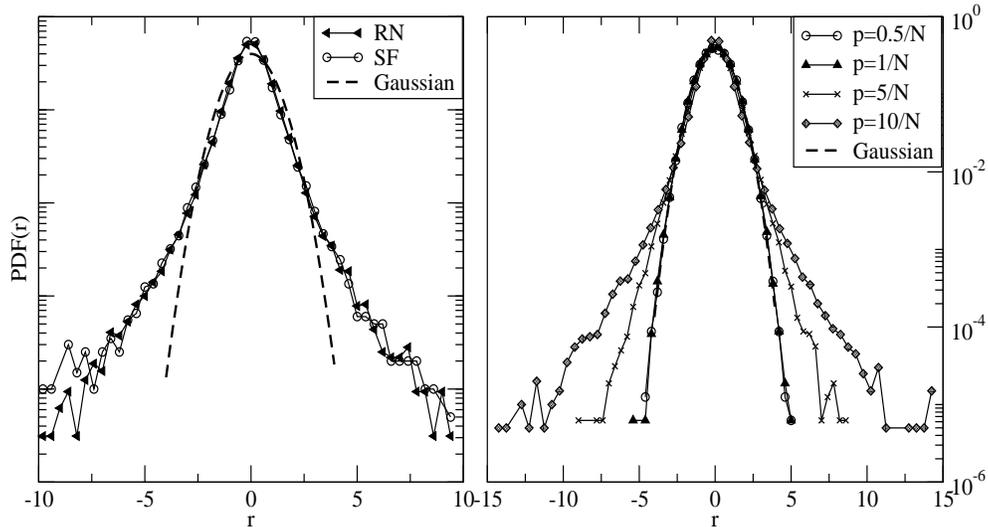}}
\caption{ Left: comparison between the PDFs of our model obtained on a SF network
and on a RN with number of nodes $N=10^4$ and average links per node $\bar{k}=10$.
 For the SF network the parameters used are $m=m_0=5$ for the links of each new node 
and $\theta=0.9$ for the clustering probability while for the RN $p=10/N$.   
From a statistical point of view the characteristic features of the PDFs have their
origin in the model dynamics as opposed to the fine features of the network.
Right: Dependence of the opinion fluctuations on the parameter $p$
on a RN. The parameters used for the dynamics are $a=1.8$ and $\kappa=a$
for the group and global opinion response respectively.}
\label{pdf_double}
\end{figure}
%------------------------------------------------------------------------------
%
The results, shown in Fig.~\ref{pdf_double} (left) demonstrate how the
dynamics of the two  systems are largely equivalent under the adopted
constrains.  In Fig.~\ref{pdf_double} (right) we also show the
dependence of the dynamics  on the parameter $p$ for the RN.  At the
critical threshold, that is the value of $p$ for which a giant cluster
appears, there is still no trace of turbulent-like activity giving
rise to fat tails.  Yet,  in this case each agent has, on average,
just one link and there cannot be any small world properties.

These results confirm that the critical topological characteristic
leading to herding behaviour in the framework of stochastic  opinion
formation is the presence of mean field effects enhanced by
small-world structure.  The more information (links) that an agent
has, the more likely it is for him/her to have an opinion  in accord
with other agents.

In the next section we extend our model in order to include indecision
in the process of opinion formation.

\section{The Influence of Indecision}
\label{indecision}

We now extend our model in order to include the concept of indecision.
 In practice a certain agent $i$, at a time step $t$, may take neither
 of the two possible decisions, $\sigma_i= \pm 1$, but remain in a
 neutral state.  Keeping faith to the spirit of the model, we address
 this problem introducing an {\em indecision probability}, $\epsilon$:
 that is the probability to find, at each time step, a certain agent
 undecided.  This is equivalent to introducing time dependent failures
 in the structure of the network by setting $\sigma= 0$.

\begin{figure}
\vspace{1cm}
\centerline{\epsfig{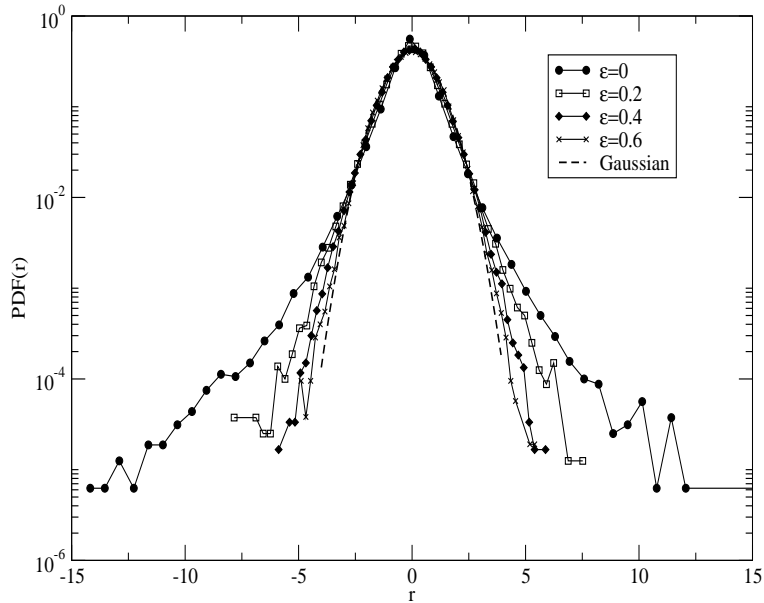}}
\caption{ Transition from coherent bahaviour, indecision probability 
$\epsilon=0$, to noise
using a random selection for the inactive agents. For $\epsilon \approx 0.6$ we
reach a noise-like behaviour. The parameters used in the simulation are:
$N=10^4$ nodes, $\theta=0.9$ for the clustering probability, $m=m_0=5$ for the
links of each new node, $a=1.8$ and $\kappa=a$ for the group and global opinion response
respectively.}
\label{pdf_rand}
\end{figure}

Focusing on the turbulent-like regime, the shape of the PDF in the
opinion fluctuations changes according to different concentrations of
undecided persons.  The results of the simulations,  in
Fig.~\ref{pdf_rand}, show how the dynamics of the model move from an
intermittent state for $\epsilon=0$ toward a noise state for $\epsilon
\approx 0.6$.  The convergence to a Gaussian distribution is obtained
only for quite  high concentrations of undecided agents at about 60\%.
The robustness of the turbulent-like behaviour is related to the
intrinsic robustness of SF networks against random
failures~\cite{tolerance}.  In fact, because there is a large absolute
number of poorly connected nodes, related to the power law shape of
$P(k)$, the probability of setting one of them to inactive is much
higher compared to the ``hubs'' that are relatively rare.

%In fact, because the large absolute number of poorly connected nodes,
% due to the power law shape of $P(k)$, the probability
%of setting to inactive one of them is much higher compared to the ``hubs''
%that are relatively rare. 

We can claim that, in large social networks governed by stochastic reactions
in their elements, large fluctuations
in the average opinion can appear even in the case in which a large part of
the network is actually ``inactive'' provided that the structure is scale
free and the indecision is randomly distributed.
The existence of large hubs provides for the survival of extended sub-networks
in which synchronization can give rise to coherent events. The structure
of the network itself supplies the random indecision. 

%These results 
%confirm those found for percolating clusters network in Ref.~\cite{bartolozzi04},
%where the synchronization of just a few clusters was responsible
%for crashes.

%--------------------------------Figure 7--------------------------------------
%------------------------------------------------------------------------------

\begin{figure}
\vspace{1cm}
\centerline{\epsfig{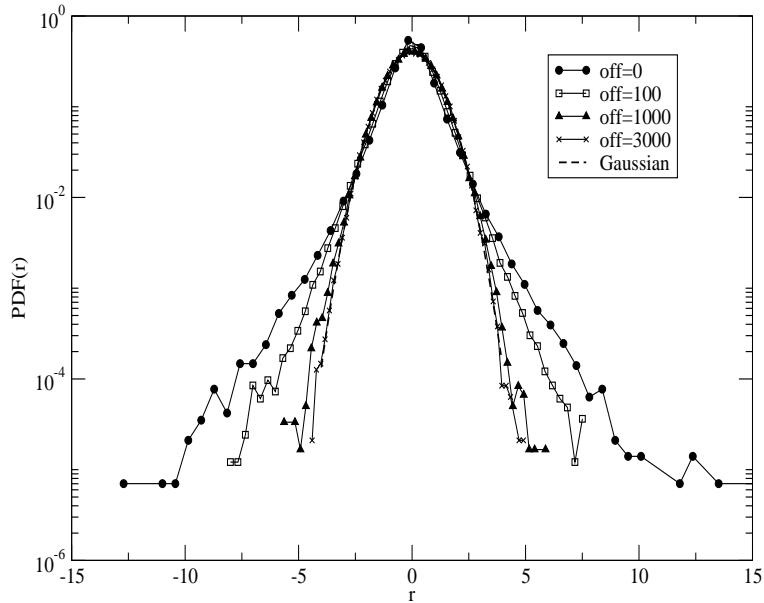}}
\caption{ In this simulation we progressively turn off the largest 
hubs in the network. Once we have turned off about the 10\% of agents, $N=10^4$,
the coherence in opinion formation disappears. The parameters used
in the simulations are the same as in  Fig.~\ref{pdf_rand}.}
\label{pdf_aim}
\end{figure}

Now we address the question of how the dynamics may change if we do
not choose randomly the inactive nodes but we target the nodes having
the most links.  What we do in practice is to sort the nodes according
to their number of links and then deactivate the nodes having the largest
number of links in decreasing order.  Fig.~\ref{pdf_aim} illustrates
how the fragmentation process is much faster and the noise regime is
reached already when only the 10\% of the hubs are deactivated.  As
emphasized in Ref.~\cite{tolerance}, the hubs have a great importance
in the structural properties of SF networks and specifically targeting
these nodes can lead to sudden isolation of a large fraction of the
nodes of the network.

%--------------------------------Figure 8--------------------------------------
%------------------------------------------------------------------------------

\section{Agent Induced Indecision: the Three State Model}
\label{3states}

In the previous section we introduced random and targeted failures in
order to study the response of the system to changes in the network
topology.  In a real social network the reason behind the indecision
of a person follows much more complex rules and can depend on
different factors as, for example, unsatisfactory information
obtained by his/her sources.  As seen from Eq.~(\ref{field}), the
opinion of each agent depends on the  poll of his/her network links. 
 Suppose now that the agent $i$ has $\tilde{N}_i$ neighbours where
$\tilde{N}_i/2$ of these share the opinion +1 while the remaining
$\tilde{N}_i/2$ share the opposite opinion. In this case, unless we
give specific weights to each node, the agent $i$ will not have an
easy task in choosing one of the two possible positions because of a
lack of popular consensus.  Based on this idea derived from  common
sense, we can extend our two state model by introducing an {\em
induced indecision probability}, $\mu$,  dependent on the information
available to the agents at each time step.  In particular we define
the global opinion of the neighbours of the $i$th node as
$s_i(t)=\sum_{j=1}^{\tilde{N}_i}\sigma_j(t)$ and the
indecision probability for the $i$th node at time $t$
\begin{equation}
\mu_{i}(s,t)=c_i \  e^{-s_{i}^{2}(t)/2\varsigma},
\label{mu}
\end{equation}
where the indecision probability width, $\varsigma$, is a parameter of the model and 
$c_i$  a normalization constant that depends just on
the structure of the network. It  calculated at the beginning of the
simulation by imposing $\sum_{s=-\tilde{N}_i}^{\tilde{N}_i} \mu_i(s,0)=1$, i.e.
the sum of the indecision probabilities over all possible global opinions to be one.
The model of Eq.~(\ref{mu}) assumes a Gaussian probability, centered in $s_i=0$,
 for the distribution of indecision of the $i$th agent. That is, the probability
of having this agent in a state with $\sigma_i=0$ is 
greater when there is not a large agreement
in the opinion of the his/her sources.  

\begin{figure}
\vspace{1cm}
\centerline{\epsfig{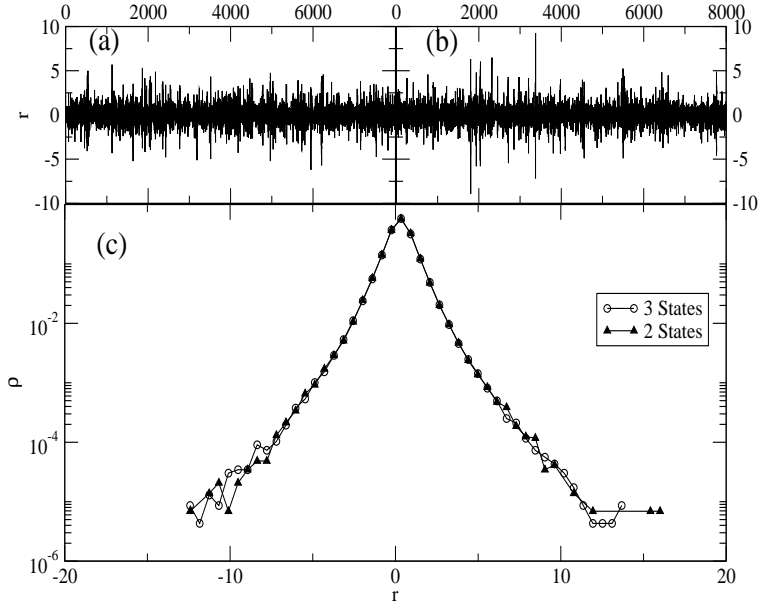}}
\caption{ (a) A window of the normalized time series generated 
by the two-state model with parameters 
$N=10^4$ nodes, $\theta=0.9$ for the clustering probability, $m=m_0=5$ for the
links of each new node, $a=1.8$ and $\kappa=a$ for the group and global opinion response
respectively. 
(b) Window of the normalized time series generated by 
the three states model with the same parameters
as (a) and indecision probability width
 $\varsigma=1$. (c) Comparison between the PDFs generated by the
two and three-state models with the aforementioned parameters obtained
over 50 realizations of the SF network. No relevant differences can be observed.}
\label{comp_3s}
\end{figure}

The analysis of the time series generated by the three state model 
does not present any relevant difference if compared with the
two state model with the same parameters, Fig.~\ref{comp_3s}.

%--------------------------------Figure 9--------------------------------------
%------------------------------------------------------------------------------

\begin{figure}
\vspace{1cm}
\centerline{\epsfig{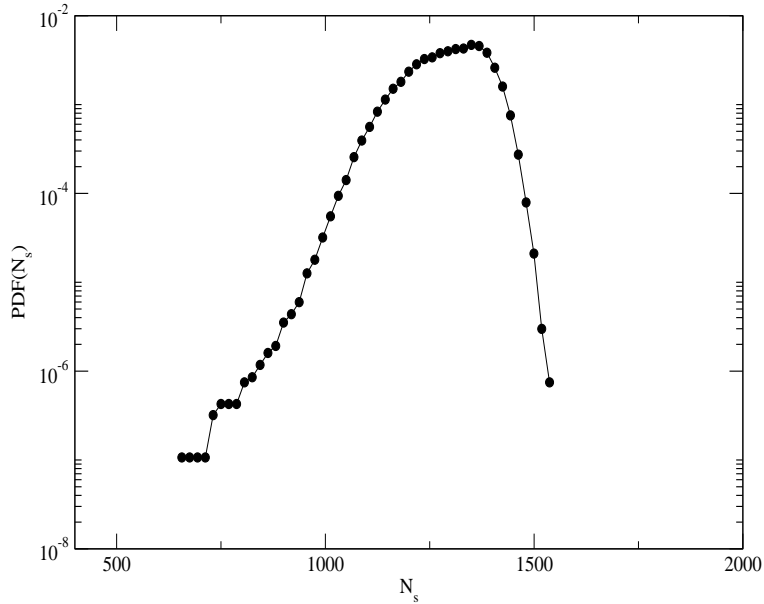}}
\caption{ PDF of the number of inactive agents, $\sigma_i(t)=0$,
during the simulation of the three state model. The parameters used are the same as 
used in Fig.~\ref{comp_3s}.}
\label{pdf_inac}
\end{figure}

We also plot the PDF for the number of inactive agents, $N_s(t)$,
during the simulation, Fig.~\ref{pdf_inac}.  It is interesting to notice
how this distribution is not Gaussian distributed around the average but
it is skewed on one side. Moreover, only a small fraction 
of agents is undecided,
of the order of 10/15 \%. This is consistent with the observation that in 
opinion polls most of the participants actually indicate an opinion.

%--------------------------------Figure 10--------------------------------------
%------------------------------------------------------------------------------

\section{Possible Application: Opinion Formation and the Stock Market}
\label{dj}

The model for opinion formation discussed thus far can be tested
against the best known real social network: the stock market.  The
main idea is to compare our results with some {\em stylized facts}
concerning the price time series, $P(t)$ and, in particular, with the
properties of the  logarithm of the price fluctuations, or {\em
returns}, $R(t)=\ln P(t+1)-\ln P(t)$. In fact some characteristic
features are independent of  the particular market and can be
considered as universal~\cite{mantegna}.  Moreover the returns show an
intermittent behaviour, reminiscent of hydrodynamic
turbulence~\cite{mantegna,mantegna95,ghashghaie96,mantegna97}, also
characterized by power law tails in the PDF.  In this case the large
coherent events are related to  crashes or  other  anomalous
variations of price.

\begin{figure}
\vspace{1cm}
\centerline{\epsfig{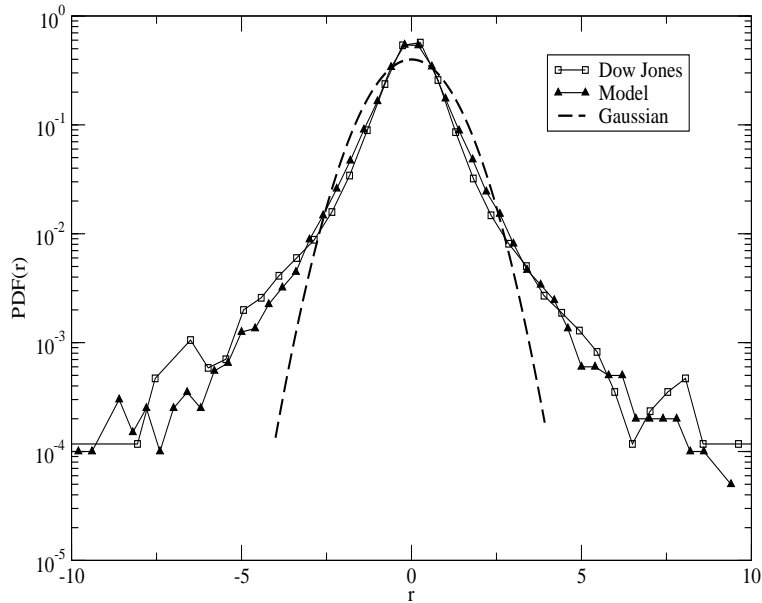}}
\caption{ Comparison between the PDF of our model and the time series of
the Dow Jones index from 13/1/1930 to 13/4/2004. The parameters of the model
used to reproduce the PDF in the plot are: 
$N=10^4$ nodes, $\theta=0.9$ for the clustering probability,
 $m=m_0=5$ for links of each new node, $a=1.8$ and $\kappa=a$ 
for the group and global opinion response respectively.
 A Gaussian is also superimposed in order to emphasize the fat tails.}
\label{pdf_dj}
\end{figure}

 If we assume that the variation of price is directly proportional to
 changes in demand and supply,
\begin{equation}
 \frac{dP}{dt} \propto c_p \  P,
\end{equation} 
where $c_p$ is
proportional  to the average opinion, $r(t)$, then the returns are
proportional to the average opinion $R(t) \approx
r(t)$.  Using this assumption, we compare the time series of average
opinion generated by the two state model against the   time series of daily
closures of the Dow Jones  index. The data set spans the range 13/1/1930
to 13/4/2004 for a total of 18645 samples.  In Fig.~\ref{pdf_dj},
a comparison between the two PDFs is shown.  The similarities between
the model and the Dow Jones is remarkable.  Both distributions
have a {\em leptokurtic} shape and, in particular, they are
described by power law tails, expressing the turbulent-like
dynamics of the time 
series\footnote{The problem of the actual shape of the PDF for
the stock market returns is still a matter of debate in the econophysics 
community~\cite{mantegna,market_dist}. A solution to this problem
would be of a great  interest, especially for the practical application
of option pricing.}. Note that, in contrast to the self-organized model
for stock market dynamics proposed by Bak {\em et al.}~\cite{bak97}, here 
the price feedback 
is not an essential ingredient for the reproduction of the correct shape 
of the distribution. Rather it is  the herding behaviour that plays the main role,
 as observed from Fig.~\ref{pdf_k}.

%--------------------------------Figure 11--------------------------------------
%------------------------------------------------------------------------------

\begin{figure}
\centerline{\epsfig{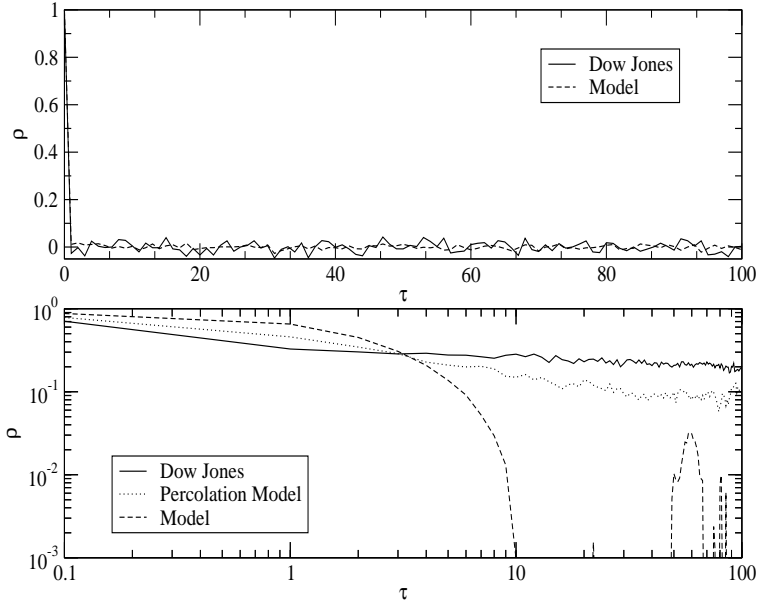}}
\caption{ Autocorrelation functions for the fluctuations $r$ (top) and
the  volatility $\nu$ (bottom). The parameters used to produce the
analyzed set are: $N=10^4$ nodes, $\theta=0.9$ for the clustering probability,
 $m=m_0=5$ for the links of each new
 node, $a=1.8$ and $\kappa=a$ for the group and global opinion response
respectively.}
\label{correlation}
\end{figure}

The similarities between the artificial time series generated by the
virtual social network and the stock market extend beyond the fat tails 
in of PDF of the fluctuations  to temporal correlations.  
It is well known that the stock market
returns have negligible correlations  on daily intervals while the
{\em volatility}, $\nu$,  defined as their absolute value, have a slow
power law decrease as a function of the time lag. This phenomenon is known as
{\em volatility clustering}~\cite{mantegna}.  In order to make a
comparison with our model  we make use of the autocorrelation
function, $\rho$.  For a time series of $L$ samples, $x_i$ for
$i=1,...,L$, this is defined as
\begin{equation}
\rho(\tau)=\frac{\sum_{j=1}^{L-\tau}(x_{j}-\bar{x})(x_{j+\tau}-\bar{x})}
{\sum_{j=1}^{L-\tau}(x_{j}-\bar{x})^2},
\label{autoc_func}
\end{equation}
where $\tau$ is a time delay and $\bar{x}$ represents the average over
the period under consideration.    The autocorrelation has been
computed both for the returns and for the volatility.  While the time
series of returns  generated by the model and the Dow Jones index have
an equivalent behaviour, Fig.~\ref{correlation} (top), the same
similarities do not hold for the volatility, Fig.~\ref{correlation}
(bottom). We observe a qualitatively different correlation: while for
the market we observe a power law behaviour, the memory in the time
series generated by the model  decays exponentially like a short-range
correlated random processes~\cite{mantegna}.   This second point
illustrates how non-trivial memory effects in the stock market  cannot
be taken into account by a simple heath bath dynamics.

%--------------------------------Figure 12--------------------------------------
%------------------------------------------------------------------------------

In Fig.~\ref{correlation} (bottom) we also reproduce the
autocorrelation function for the model presented in
Ref.~\cite{bartolozzi04}.  In this model an heat bath dynamics,
similar to the one used in the present simulations, is applied to
dynamical percolation clusters, used as a paradigm for agents
aggregation. The temporal evolution of the clusters, which size
follows a power law distribution,  is related to a forest-fire
dynamics in which some potential traders are attracted in the market by other
already active traders while, at the same time, some of them may temporally
quit the trading. Large fluctuations in the price changes are due to the
synchronization in the of the larger clusters in the market at a
particular time.  The main qualitative difference between this model
and the one presented so far is that the former presents a decay rate
much closer to that  of the real market. At this point it is important
to underline that the main difference between the two models is
related to the network dynamics. While in the present simulation the
network is fixed, in Ref.~\cite{bartolozzi04} the interaction between
agents are time dependent and localized in separate clusters.  We can
argue that the dynamics of the networks and, in particular,  the
clustering of agents in different sub-networks can  play an important
role in the correlation properties of the stock market volatility. In
reality, this fact appears quite natural if we use the autocorrelation
function, defined in Eq.~(\ref{autoc_func}),  in order to estimate the
degree of memory in a process.  If, for example, the variable under
investigation is the sum  over many independent Markovian processes,
as in Ref.~\cite{bartolozzi04}, then the resulting autocorrelation is
given the convolution of the common exponential decay, $\propto
e^{-\beta \tau}$, with the distribution of the decaying rates,
$g(\beta)$,
\begin{equation}
 \rho(\tau) \propto \int_{0}^{\infty} g(\beta) e^{-\beta \tau}d\beta.
\end{equation}
 According to the shape of this distribution, the observed macroscopic
variable can show a behaviour characteristic of a long memory
processes, like the $1/f$ Fourier spectrum~\cite{ziel50}. Power law tails
in the probability distribution function, 
$\rho(\tau)= \tau^{-\gamma}$, are produced from the distribution 
$g(\beta)=\Gamma(\gamma)^{-1}\beta^{\gamma-1}$, 
where $\Gamma$ is the gamma function and $\gamma$ a generic 
real exponent~\cite{sornette_cf}.
  This fact strengthens the idea that
the stock market  is organized in a hierarchy of sub-networks where
each of them can be considered, from a physical point of view, at
local equilibrium. For time periods shorter than the typical time
scale necessary for the networks to evolve, the only link between the
sub-systems composing the market is the feedback coming from the price
history.  This idea is closely related to the concept
of {\em subordination} used in probability theory~\cite{feller}. The 
superposition of distributions, as a possible explanation of fat-tailed processes,
 has been proposed recently by Beck~\cite{beck} in the context of
hydrodynamic turbulence  and then extended also to other
systems~\cite{beck05} including the stock market~\cite{ss_stock}.

\section{Multifractal Analysis}
\label{mf_analysis}

Financial time series present an
inherent {\em multifractality}~\cite{feder}.  In the past few years
the work of many
authors~\cite{rodrigues01,gorski02,auloos02,dimatteo03,bartolozzi04} has been
addressed to the characterization of the multifractal properties of
financial time series, and nowadays multifractality can be considered
as a stylized fact.  In order to study the multifractal properties of
our model we use the {\em generalized Hurst
exponent}~\cite{mandelbrot}, $H(q)$, derived via the $q-$order
structure function,
\begin{equation}
S_{q}(\tau)= \langle |x(t+\tau)-x(t)|^{q}{\rangle_{T}} \propto
\tau^{qH(q)},
\label{structure}
\end{equation} 
where $x(t)$ is a stochastic variable over a  time interval $T$ and
the time delay, $\tau$.  The generalized Hurst exponent, defined in
Eq.~(\ref{structure}),  is an extension of the Hurst exponent, $H$,
introduced in the context of reservoir control on the Nile river dam
project,  around 1907~\cite{feder,hurst51}. This technique   provides
a sensitive method for revealing long-term  correlations in random
processes.  If $H(q)=H$ for every $q$, the process is said to be
monofractal and $H$ is equivalent to the original definition of the
Hurst exponent. This is the  case of simple Brownian motion or
fractional Brownian motion.

\begin{figure}
\centerline{\epsfig{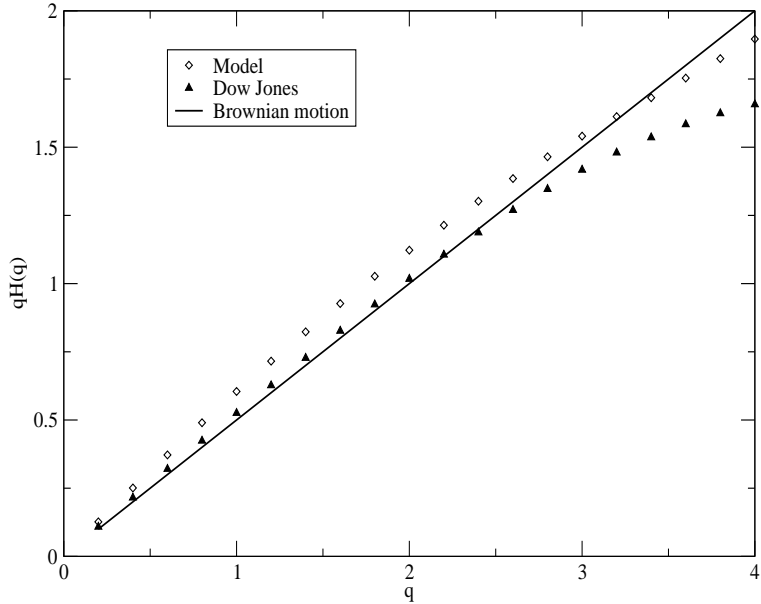}}
\caption{ Structure function exponents for the Dow Jones index and our
model. A deviation from a linear behaviour is evident. The
hypothetical  spectrum of a 1D Brownian motion is also shown for
comparison.}
\label{multifractal}
\end{figure}

If the spectrum of $H(q)$ is not constant with $q$  the process is
said to be multifractal.  From the definition (\ref{structure}) it is
easy to see that  the function $H(1)$ is related to the scaling
properties of the volatility.  By analogy with the classical Hurst
analysis, a phenomenon is said to be persistent if $H(1)>1/2$ and
antipersistent if $H(1)<1/2$.  For uncorrelated increments, as in
Brownian motion, $H(1)=1/2$.  In Fig.~\ref{multifractal} a comparison
is shown between the multifractal spectra of the  model and the Dow
Jones index obtained from the price time series.  It is clear that
both processes have a multifractal structure and the price
fluctuations cannot be associated  with a simple random walk as in the
classical  {\em efficient market hypothesis}~\cite{bachelier00}.

%--------------------------------Figure 13--------------------------------------
%------------------------------------------------------------------------------

\section{Discussions and Conclusions}

In the present work we have introduced a two state model  of opinion in
order to simulate the complex dynamics of opinion formation in a group
of individuals. The decision updating is governed by a stochastic heat-bath 
dynamics that mimics the reaction of each person to his/her
specific sources of information as governed by the network
neighbours and to the average opinion of the
whole group. Particular emphasis has been given to the topology of
the interactions between agents, where a Barab$\acute{a}$si-Albert scale-free
network has been used to simulate the links between them. The choice
of this particular network is motivated by a series of recent
studies on social  aggregation~\cite{albert02, dorogovtsev02} but, as we have shown 
in Sec.~\ref{numerical}, its use is not essential for the appearance of
 coherent events.
As in other studies~\cite{kaizoji,krawiecki02,bartolozzi04}, we find a range in
the parameter space in which the fluctuations of opinion have a
non-trivial turbulent-like dynamics.  The results of the simulations
show that the most important factor determining the appearance of
large fluctuations, is the synchronization of large parts of the
network. As discussed in  Sec.~\ref{numerical}, this feature plays an important
role even in the case in which the  personal opinion is
relatively strong.  As a consequence large
coherent events are more likely to occur when the  average number of
links per agent is larger. 

 The topology of the interactions also plays a
key part in the dynamics of the model. In fact, introducing inactive agents
and spreading the undecided agents  randomly on the network, does not 
spoil the turbulent-like state even for high
concentrations of ``gaps'', up to approximately 60\% of agents. This is a
consequence of the implicit robustness of SF networks against random
failures.  If instead of selecting randomly the undecided individuals
we aim directly to the  ``hubs'' of the network 
then the situation changes.  In this
case the network is disaggregate, composed of very small sub-networks  and
isolated nodes. Synchronization cannot significantly effect the
resulting global opinion and the time series approximates Gaussian noise. 
We also introduce, in Sec.~\ref{3states}, 
a three state model. While the dynamics
does not significantly differ from the two state 
model, we find a persistence of 
opinion with a sharp upper limit in the number of undecided agents.
In Sec.~\ref{dj} we test the results of the simulations against a time
series of daily closures for the Dow Jones index.  The stock market,
in fact, can be considered as the most studied  network of social
interactions. The results show a very good agreement with some
stylized facts of the financial market like the broad tails in the
PDFs, temporal correlations and a multifractal spectrum. 
We also notice an interesting discrepancy in the autocorrelation function
for the volatility. Comparing the present results with those obtained
 in Ref.~\cite{bartolozzi04}, we conjecture
that the persistence in the volatility memory can be explained by
considering the market as constituted by sub-systems at local
equilibrium and weakly interacting with each other. It will be interesting 
to explore this conjecture in a quantitative manner in a further investigation.

% As a final
%remark we would like to mention how the present results confirm the
%conclusions in Ref.~\cite{bartolozzi04} where the market crashes were
%related to the synchronization of large clusters of agents. In the
%present case the synchronization is related to  large branches of the
%network, often linked by ``hubs''.  These ``hubs''  play a
%fundamental role in the stock market contest since they are the source
%of information for many agents.  The ``mood'' of these well connected 
%special agents can drive the market from an ``efficient'' dynamics, characterized by
%Brownian fluctuations, toward a high volatility state where large
%fluctuations become dominant, implying a larger risk factor for
%investors.

\begin{acknowledgments}
This work was supported by the Australian Research Council.
\end{acknowledgments}

\end{document}